**Mpemba Effect, Shechtman's Quasicrystals and Students' Exploring Activities**

Authors:

MAREK BALÁŽOVIČ

Hronská 3, Zvolen, Slovakia,
balazovicm@gmail.com,

Constantine the Philosopher University in Nitra, Slovakia

BORIS TOMÁŠIK

Tajovského 40, Banská Bystrica, Slovakia

boris.tomasik@umb.sk

Matej Bel University in Banská Bystrica, Slovakia

**Abstract**
In the 1960s, Tanzanian student Erasto Mpemba and his teacher published an article with the title "Cool" in the journal Physics Education (Mpemba, E. B. – Osborne, D. G.: Cool?. In: Physics Education, vol.4, 1969, pp. 172-175.). In this article they claimed that hot water freezes faster than cold water. The article raised not only a wave of discussions, and other articles about this topic, but also a whole series of new experiments, which should verify this apparent thermodynamic absurdity and find an adequate explanation. Here we give a review with references to explanations and we bring some proposals for experimental student work in this area. We introduce Mpemba Effect not only as a paradoxical physics phenomenon, but we shall present a strong educational message that the Mpemba story brings to the teachers and their students. This message also creates a bridge between this phenomenon and the discovery for which the 2011 Nobel Prize in Chemistry was awarded. It leads to critical adoption of traditional knowledge and encourages resilience in investigative exploration of new things.

**Introduction**
It is not hard to give positive answer to the question if hot water can freeze faster than cold water. For example, if we let a drop of hot water and a bucket of cold water cool in the same freezer, it is clear that a drop of hot water freezes sooner. Likewise, hot water in the freezer freezes sooner than cold water which we have poured out and let it cool just in a warm summer night. So the question shall be formulated more precisely in the following example: If we take two equal water samples differing only by their temperatures, place them in identical containers that are left to cool in the same way, is it possible that warmer water freezes first?

Mpemba was an ordinary African high school student in the 1960s. An interesting freezing phenomenon of hot water was first observed in connection with making ice cream. He noticed that when he placed hot milk mixed with sugar (ice cream mixture) into the freezer, it would freeze in a shorter time compared to when he had allowed the mixture to cool before. Mpemba asked his physics teacher about the cause of this, but the teacher responded to him that he had made a mistake and that this phenomenon cannot have occured. Later, Mpemba found that faster freezing of hot milk was known among ice cream producers, who prepared their products more quickly by this method. Therefore he started experiments with cooling again. He found that hot ice cream mixture not only freezes faster, but that it applies to water as well. Initial temperature difference was significant. Water hotter than 90 ° C freezes faster than water with initially room temperature in the same conditions.

**Didactic message**

Mpemba repeatedly demanded an explanation from his teacher. But the teacher still insisted that the phenomenon cannot have occurred. If Mpemba would have been satisfied with the claim of his teacher and become a passive recipient of educational explanations, everything would have been probably fully forgotten soon. However, he remained active in seeking answers to actual observation and his researching activities despite rejection and ridicule by his classmates and teachers. Several months later, he met with Dr. Osborne, a university physics teacher who visited Mpemba's high school. Mpemba took the opportunity and asked his unanswered question. Dr. Osborne didn't reject Mpemba's observation a priori and decided to explore it. In his laboratory experiments he also found that warmer water freezed faster in some cases. The previous observations were confirmed.

By its character, Mpemba's story closely resembles the discovery story of the Israeli Daniel Shechtman. Shechtman discovered quasicrystals for which he was awarded the 2011 Nobel Prize in Chemistry. He became a target of criticism by fellow scientists and wider professional community after the announcement of his discovery in 1982. Head of science department where he worked even put Shechtman in a position, where persisting on his quasicrystal discover meant losing his job. Shechtman, like Mpemba, insisted on his observations and failed to respond to textbook precepts. He let other crystallographers verify his observations independently. When observations were confirmed, he managed to find one scientist who was willing to publish an article about their paradoxical discovery with him. After publication of the article in 1984 [1], a new era of crystallography has begun. Both discoveries of paradoxical facts point out several aspects: teachers as well as scientists are reluctant to accept a new record, which at first glance contradicts the generally introduced concepts of observed reality. Our belief in traditions is often stronger than the new experimental evidence. New things are viewed with great skepticism, while the textbook records are rather uncritically accepted. When unexpected results are observed, students, but also teachers, and even some scientists tend not to study them in depth but correct the measured data in order to achieve the expected "correct" values. This approach not only kills the investigative spirit, but impoverishes the opportunity to explore new things, which in the history of science, has often come down an unexpected path. It is therefore necessary to highlight the relentless approach of Mpemba, Shechtman, or other serious researchers who had to face the counter thoughts of their colleagues, yet remained faithful to experimental evidence of their own work.

**Known – unknowns among the discoveries**
Paradoxical fast freezing of hot water is named after Mpemba. Mpemba effect, although without this term, however, occured in human knowledge much earlier. The first record of that is from Aristotle [2]. He describes people who accelerated the cooling process of water by leaving water to get warm in the sun before cooling. He reported also about fishermen who used hot water to water their rods, which were fixed in this way into the holes in the ice during the winter. Other scholars captured the fact that warmer water freezes faster, too. They were people like Roger Bacon [3], René Descartes [4], and Francis Bacon [5]. Cooling process of warm liquids was regularly used in ice cream and ice drinks production in India in the 18$^{th}$ century [6]. However, with the advent of modern theories of heat transfer, these observations retreated into the background in the field of scientific discussion and general knowledge.

Shechtman's discovery of quasicrystals does not have such a far-reaching history as Mpemba Effect. It is also due to the methods of examining the diffraction pattern of crystals, which are not so old. However, it is very interesting, that what helped the scientists to understand distribution of atoms in quasicrystals were medieval works of art. Mosaics located in Alhambra and in Iran from the time of the Middle Ages have similar structure like quasicrystals.

**Explanations**

Despite more than 2000 years of history, Mpemba phenomenon hasn't got its uniform explanation, which has its reasons. Since the issue of the Mpemba article, dozens of other contributions have come out which represented the explanations of the phenomenon documented by the results of their experiments. The entire mosaic of interpretations has been formed in the past forty years. Each of its parts looked like the only true one. The reason may be the fact that each of the observers studied the phenomenon in different conditions. In a seemingly simple experiment of cooling warm and cool water, there is a wide range of factors affecting the temperature decrease of water in time. They are, for example, the properties of the liquid container in which it is located, or the properties of the cooling environment. The cause of the phenomenon may be different for each case. In addition, the phenomenon does not always occur. Therefore, the first step to its investigation is finding the conditions under which it can be observed. The effort to find common explanation for the emergence of the phenomenon, applicable for all cases, is probably not a good way. We should therefore withdraw from seeking the only single correct interpretation. Here are a few parts of the interpretative puzzle.

*Evaporation.* Warmer water evaporates faster than cold water. Because of that it reduces its volume faster. Since the time needed to freezing is directly proportional to the amount of water, it can be easy to explain this phenomenon. This explanation was one of the first after the rediscovery of the phenomenon. It is discussed in more detail by G. P. Kell [7]. The relation between the speed of temperature decrease and the size of the liquid surface was dealt with J. Walker [8]. Evaporation is well applicable as the cause of the Mpemba phenomenon in cases where liquid surface is too large. However, Mpemba effect was also observed in closed containers with no evaporation.[9] In these cases, evaporation cause is unusable.

*Chemical Composition.* Heating process of water causes not only its temperature change, but also other changes, like change of the quantity of dissolved gases and solid substances contained in water. By heating water, the gases are expelled and solids reduce their concentration. The influence of gas on the cooling rates of water has been recorded by B. Wojciechovski or M. Freedman [9][10]. Smaller quantities of gases and solids in the water can increase thermal conductivity, flowing in the water or may cause a shift in the freezing point of water.

*Supercooling.* The freezing temperature of water at normal pressure is often identified with the temperature of 0 ° C. Freezing process, however, usually starts at a lower temperature. Water may remain in the liquid state sometimes even at much lower temperatures, and we call it supercooled water. One of the explanations of the phenomenon is that initially warmer water is supercooled less than initially cooler water. The first to come up with this explanation was D. Auerbach [11], later followed by M. Duffy or J. A. Chaplin [12] [13]. Different levels of supercooling may result from differences in chemical compositions and changes in the configuration of molecules into larger structures.

*Microstructure of water.* Water molecules are not completely isolated from one another and they often merge, creating more complex structures called clusters. [12] Molecules of water transform their positions into a certain characteristic position when ice is formed. Complicated clusters are not the most appropriate design units for ice forming. Just as it is easier to build a Lego-house from separate cubes than from improperly connected pieces, ice is more easily formed from separate molecules of water than from clusters. How to make separate molecules out of clusters? You just need to heat the water sufficiently. The clusters burst and better units for the construction of the house of our ice arise more easily. More clusters in cooler water can cause ice forming at lower temperatures.

*Flowing.* In warmer water there are stronger streams. The water circulates and the heat is transfered towards the walls of the container more quickly. The inertia can provide significant flow of water during its entire cooling. For hot water, this means faster cooling and the water can catch up the advantage of cool water. Flowing effect on cooling liquid was studied by E. Deeson [14] and I. Firth [15].

*Contact with the cooler.* If the containers are cooled in a freezer with ice coating, the container with hot water can melt ice under itself. Ice is not a good heat conductor. If it is melted below the jar more, the jar will be closer to the freezer. The heat will then be eliminated faster. Mpemba himself considered this the most likely interpretation for his first observation of the phenomenon.

**Student activities**

Whereas the phenomenon has no definitive explanation, it is an appropriate candidate to students' experimentation. In addition, these activities don't need any special equipment. The phenomenon doesn't occur regularly, therefore it is interesting to search conditions when it may arise. It provides a wide option of various parameters, which provides a plenty of space for the discussion of different results from similar initial conditions. In examining the various explanations, it is appropriate to be inspired by the above mentioned resources referred to in the text. In spite of this, we offer several types of exploring activities of your investigation of freezing water:

1) Make an ice cream mixture like Mpemba, cool it in your freezer and measure temperatures for different initial temperatures of mixtures.

2) Measure temperatures in the cooling and freezing process of bodies of water with the same initial temperatures, but cover the first container and not the second one.

3) Measure decreasing temperature of water in the freezer, which has the same initial temperature, but different shapes of containers.

4) Measure temperatures of water in cooling process and use water with different hardness.

5) Heat water to the same initial temperature, put it to the freezer and then make measurements of temperatures. You mix one of the samples during cooling and search differences between cooling rates of the two water bodies.

6) Make temperature measurements of water in different freezers (freezer with frost, without it, freezing outside if there are conditions, etc.)

7) Use distilled water in a clean container and monitor possible supercooling process of the liquid.

8) Find methods which could verify inertia and stronger flowing in hot water throughout the whole cooling process.

**Simple experimental set up**

Individual measurements and examinations of the explanations of the phenomenon can be realized within the school project for individuals or entire teams. Measurements improve not only the experimental skill, but also emphasize the ability of observation, careful monitoring and a detailed description of measurement conditions, which are often absent in the articles presented on this topic.

Measurements can be realized also with simple equipment. A freezer can be created by ice (snow) and salt. When salt is added to the ice, it lowers the freezing point of the ice, so even more energy has to be absorbed from the environment in order for the ice to melt. This makes the ice mixture colder than it was before. In this way we can create a homemade freezer with temperatures about -20°C.

**Some results**

We did some experiments with this salt/ice mixture. We put this mixture in a glass aquarium and we insulated it by a polystyrene layer. We placed four plastic bottles into the mixture, each filled with 0.5 litre distilled water. The thermometers were inserted in the water through the bottle caps (Figure 1).

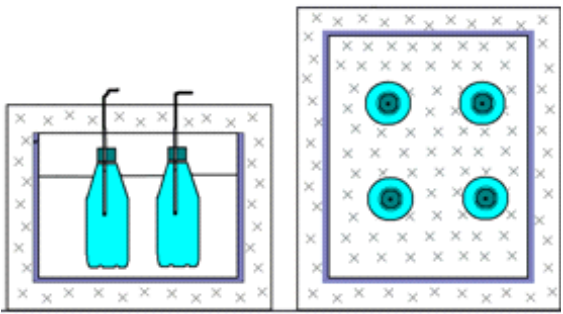

**Figure 1:** Side view (left) and top view (right) of the bottles with distilled water in ice/salt mixture placed in an insulated aquarium. Thermometers were inserted through the bottle caps.

Then, we were measuring the temperature every second and observed if water in the bottles with approximately the same initial conditions would have the same time of freezing. Temperature curves are recorded in Figure 2. We can see differing curves and freezing points of events. Water stayed in liquid form until the first crystals were created around some heterogeneity or impurity. It took different times. This was evoked by the sensitivity of measured system. In spite of very similar initial parameters, the water started its freezing process at different times.

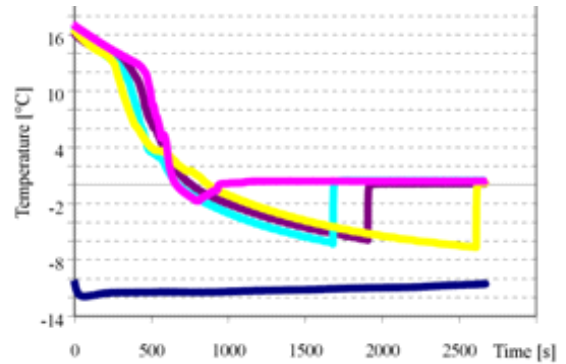

**Figure 2**: Temperature curves of water cooling in bottles from the same initial temperature. The lower curve (blue) represents the temperature of the freezer (ice/salt mixture).

**Conclusion**

It seems that crystalline water can still surprise us and not only shake our misconceptual knowledge, but also influence our attitude towards exploring the new. Mpemba effect is not a singular phenomenon in physics teaching process, which can cause problems with explanations. All teachers from time to time expierence some unfamiliar effect or unexpected observations. We can find a lot of similar non-intuitive physical effects described in publications. Let us mention an article by R. Danson [16], who did a cooling experiment and found that hot water cooled down faster in calorimeter insulated with a piece of card than hot water in uninsulated copper calorimeter.

Mpemba and Shechtman's stories tell us about our distrust and ignorance of paradoxical observations. At the same time, however, they show that it pays to believe in experimental methods, which always push our knowledge one step further.